\documentclass[final,journal,letterpaper]{IEEEtran}

\usepackage{amsmath}
\usepackage{graphicx}
\usepackage{array}
\usepackage{hyperref}
\usepackage{microtype}
\usepackage{multicol}
\usepackage[caption=false,font=footnotesize]{subfig}

\newcounter{MYtempeqncnt}

\begin{document}

\title{Performance of Opportunistic Epidemic Routing on Edge-Markovian Dynamic Graphs}

\author{John Whitbeck, Vania Conan, and Marcelo Dias de Amorim

\thanks{A poster of this work was presented at the \emph{ACM SIGCOMM Workshop on Networking, Systems, Applications on Mobile Handhelds (Mobiheld 2009)}. This version is more detailed and contains many more results.}%

\thanks{This work has been partially supported by the ANR project Crowd under contract ANR-08-VERS-006.}%

\thanks{John Whitbeck is with both UPMC Sorbonne Universit{\'e}s and Thal{\`e}s Communications, France. Email: john.whitbeck@lip6.fr. }

\thanks{Vania Conan is with Thal{\`e}s Communications, France. Email: vania.conan@fr.thalesgroup.com.}

\thanks{Marcelo Dias de Amorim is with LIP6/CNRS~-- UPMC Sorbonne Universit{\'e}s, France. Email: marcelo.amorim@lip6.fr.}}

\IEEEpubid{0000--0000/00\$00.00 ̃\copyright ̃2010 IEEE}

\markboth{IEEE Transactions on Communications, TCOM-09-0163}%
{Whitbeck \MakeLowercase{\textit{et al.}}: Performance of Opportunistic Epidemic Routing on Edge-Markovian Dynamic Graphs}

\maketitle

\begin{abstract}
Connectivity patterns in intermittently-connected mobile networks (ICMN) can be modeled as edge-Markovian dynamic graphs. We propose a new model for epidemic propagation on such graphs and calculate a closed-form expression that links the best achievable delivery ratio to common ICMN parameters such as message size, maximum tolerated delay, and link lifetime. These theoretical results are compared to those obtained by replaying a real-life contact trace.
\end{abstract}


\IEEEdisplaynotcompsoctitleabstractindextext
\IEEEpeerreviewmaketitle

\section{Introduction}
\label{sec:introduction}

Intermittently connected mobile networks (ICMN) emerge from the social processes that bring mobile devices into contact.  Due to high node mobility and frequent lack of end-to-end connectivity in such networks, message transport is usually handled in a store-and-forward fashion by delay/disruptive-tolerant network (DTN) routing protocols~\cite{dtn_fall_sigcomm}.

The topology of a real-life network of mobile devices evolves over time as links come up and down. A network's connectivity graph is defined by associating each node to a vertex and adding an edge between any pair of nodes that are currently in contact (i.e., within transmission range of each other). Successive snapshots of the evolving connectivity graph yield a \emph{dynamic graph}, i.e., a time-indexed sequence of static connectivity graphs. Their theoretical study is therefore important for understanding the underlying network dynamics.

In this paper, we propose a new Markovian model for flooding on edge-Markovian dynamic graphs~\cite{Clementi08}. Unlike previous work on asymptotic behavior~\cite{Clementi08}, our approach assumes source-destination pairs for messages, a finite number of nodes, as well as finite link capacities and message sizes. Our main contribution is a closed-form expression of the bundle\footnote{Bundles are \emph{message aggregates}. They can contain anything from a message fragment to several messages~\cite{dtn_fall_sigcomm}.} delivery ratio as a function of bundle size, maximum tolerated delay, and the dynamics of the underlying edge-Markovian dynamic graph. Using this model, we show that the message delivery ratio increases for smaller bundles, but that the achieved gain is bounded and only significant when the constraints on message delivery delay are tight. Finally, we compare our model's predictions to results from a real-life connectivity trace obtained in a rollerblading tour.

In Section~\ref{sec:markovian_graphs}, we briefly describe the edge-Markovian dynamic graph model. We then calculate the delivery ratio for epidemic routing in Section~\ref{sec:message_size}, before discussing the the impact of bundle size on delivery ratio in Section~\ref{sec:discussion}. We then compare these theoretical insights to results from a real data set, the Rollernet experiment, in Section~\ref{sec:experimental}.

\section{Edge-Markovian dynamic graphs}
\label{sec:markovian_graphs}

\IEEEpubidadjcol

In the rest of this paper, in both the theoretical and experimental parts, we represent the ad hoc network formed by $N$ mobile nodes as a connectivity graph that evolves in discrete time. Depending on the context, we will use the terms vertex (resp. edge) and node (resp. link) interchangeably. The time step $\tau$ is equal to the shortest contact or inter-contact time. In a real-life trace, $\tau$ is equal to the neighborhood scanning sampling period. Edges come up or down at the beginning of each time step, but the topology then remains static until the next time step.

Previous work on dynamic graphs focused on graphs with increasing numbers of vertices or edges~\cite{GOT}, but did not account for node mobility and/or link instability. More recently, Chaintreau \emph{et al.} used simple sequences of uniform random graphs for modeling random temporal graphs in order to analyze the diameter of opportunistic mobile networks~\cite{chaintreau_diam}. Pellegrini \emph{et al.} explored the notion of \emph{connectivity over time} but this approach loses all information about the order in which contact opportunities appear~\cite{pellegrini07}. Unfortunately, none of these models capture the strong correlation between successive connectivity graphs.

Edge-Markovian dynamic random graphs were first introduced by Clementi \textit{et al.} as a generalization of time-independent dynamic random graphs to capture the strong dependence between the existence of an edge at a given time step and its existence at the previous time step~\cite{Clementi08}. While such a model may be used to  study a wide variety of dynamic graphs, in this paper, we will focus on its application to ICMNs.

Dynamic random graph based models, including the one in this paper, have an exponential (or geometric) inter-contact time distribution. In real-life datasets this may not always be the case. Indeed, when the underlying social dynamics are strong, the inter-contact distribution follows a power law~\cite{chaintreau_mobility}. However, in different scenarios, it may follow an exponential law~\cite{lenders08}. Interestingly, the inter-contact distribution of any mobility model in a bounded domain necessarily exhibits an exponential cutoff~\cite{cai07}.

In an edge-Markovian dynamic graph with $N$ vertices, each edge is considered independently and can be in one of two states: either $\uparrow$ or $\downarrow$. Let $p_\uparrow$ (resp. $p_\downarrow$) be the probability of transitioning to the $\uparrow$ (resp. $\downarrow$) state. The transition matrix for each edge is therefore
{\small
\begin{equation}
M = \left( 
\begin{array}{c|cc}
& \downarrow & \uparrow \\
\hline
\downarrow & 1-p_\uparrow & p_\uparrow \\ 
\uparrow & p_\downarrow & 1-p_\downarrow
\end{array}
\right).
\label{eqn:edge_matrix}
\end{equation}
}

The contact ($T_\uparrow$) and inter-contact ($T_\downarrow$) times are distributed geometrically and their expected values are $E(T_\uparrow) = \frac{\tau}{p_\downarrow}$ and $E(T_\downarrow) = \frac{\tau}{p_\uparrow}$. Indeed, the number of time steps required to leave the $\uparrow$ (resp. $\downarrow$) state is the number of trials needed to get one success in a Bernoulli process with probability $p_\downarrow$ (resp. $p_\uparrow$). Let $\pi_\uparrow$ (resp. $\pi_\downarrow$) be the stationary probability of being in state $\uparrow$ (resp. $\downarrow$). We have $\pi_\uparrow = \frac{p_\uparrow}{p_\uparrow+p_\downarrow}$ and $\pi_\downarrow = \frac{p_\downarrow}{p_\uparrow+p_\downarrow}$. Finally, the average node degree is $(N-1)\pi_\uparrow$.

\section{Tuning message size to meet delay constraints}
\label{sec:message_size}

\begin{figure*}[!t]
\normalsize
\setcounter{MYtempeqncnt}{\value{equation}}
\setcounter{equation}{4}
\begin{equation}
\mathbf{T} = \left(
\begin{array}{c|ccccc}
        & $Init$ & $(1,0)$ & $(1,1)$ & $(2,0)$ & $Succ$ \\
\hline
$Init$  & 0 & \pi_\downarrow^2 & \pi_\downarrow \pi_\uparrow & 0 & \pi_\uparrow \\
$(1,0)$ & 0 & (1-p_\uparrow)^2 & (1-p_\uparrow) p_\uparrow & 0 & p_\uparrow \\
$(1,1)$ & 0 & 0 & 0 & \pi_\downarrow (1-p_\uparrow) & 1-\pi_\downarrow (1-p_\uparrow) \\
$(2,0)$ & 0 & 0 & 0 & (1-p_\uparrow)^2 & 1-(1-p_\uparrow)^2 \\ 
$Succ$  & 0 & 0 & 0 & 0 & 1 
\end{array}
\right)
\label{eqn:example}
\end{equation}
\setcounter{equation}{\value{MYtempeqncnt}}
\hrulefill
\vspace*{4pt}
\end{figure*}

\subsection{Preliminaries}
\label{subsec:preliminaries}

We assume that, when up, all links have equal capacity $\phi$ and thus can transport the same quantity $\phi \tau$ of information during one time step. We refer to $\phi \tau$ as the \emph{link size}. Small values of $\tau$ therefore mean that the network topology's characteristic evolution time is short and thus only small amounts of information may be transmitted over a link during one time step. We define the \textit{bundle size} as numerically proportional to $\tau$: $\alpha \phi \tau$. By abuse of language, however, we will simply refer to $\alpha$ as the bundle size. For example, a bundle of size $2$ ($\alpha=2$) is only able to traverse links that last for more than $2$ time steps, whereas a bundle of size $0.5$ is able to traverse two links during each time step.  Furthermore, each bundle can only tolerate a certain maximum delay. We note $d$ the maximum number of time steps, beyond which a delivery is considered to have failed. By abuse of language, we will simply refer to $d$ as the maximum delay.

Epidemic routing was one of the first methods proposed for dealing with intermittent connectivity in mobile ad hoc networks~\cite{Vahdat00epidemicrouting}. Each message is flooded into the entire network. Upon meeting, two nodes first exchange message vectors describing which messages they currently hold, before requesting from one another copies of the messages they do not yet have. Following the epidemic analogy, a node is said to be \textit{infected} by a message upon receiving a copy of it~\cite{Zhang20072867}. Epidemic routing is particularly useful for theoretical purposes, since its delivery ratio is also that of the optimal single-copy time-space routing protocol.

Our goal is to calculate the delivery ratio of a bundle using epidemic routing. To be successful, delivery has to occur without exceeding the maximum allowed delay. To this end, we introduce a new Markovian model for epidemic propagation on the edge-Markovian dynamic graph of the previous Section.

For the sake of simplicity, the model will first be described for $\alpha = 1$. In Sections~\ref{subsec:p_smaller} and~\ref{subsec:p_greater}, we will respectively describe how to adapt the previous model when the bundles are smaller ($\alpha < 1$) and larger ($\alpha > 1$) than the link size.

\subsection{Bundles fit in a time slot \texorpdfstring{($\alpha = 1$)}{}}
\label{subsec:epi_model}

Source $a$ wishes to transmit a bundle to destination $b$ using epidemic routing. Edges change states at the beginning of each time step. During one time step, an infected node infects all of its direct uninfected neighbors, and only those, since the bundle size is $1$ and bundles can therefore only perform one hop per time step. Let $V$ be the set of the nodes in the network. After $k$ time steps, nodes other than $b$ fall in one of three disjoint sets:

\begin{itemize}
   \item Those that have \emph{just} been infected: $J_k$.
   \item Those that have been infected at time step $k-1$ or before: $I_{k}$.
   \item Those that have not yet been infected: $S_k = V \setminus (I_k \cup J_k \cup \{b\})$.
\end{itemize}

This distinction is necessary to determine who can be infected at time step $k+1$. Indeed, if a node belongs to $I_k$, then all its neighbors at the end of time step $k$ are in $I_k \cup J_k$. It can only infect new nodes if an edge to a clean node in $S_k$ comes up at time step $k+1$. However, a node in $J_k$ may have edges to some clean neighbors in $S_k$ which may become infected at time step $k+1$ if the edge remains up.

In this paper, we are only interested in the probability that $b$ receives a copy in at most $d$ time steps. In this case, the only information necessary to characterize the state of the epidemic is the number of nodes $i$ and $j$ in $I_k$ and $J_k$, respectively. The delivery ratio can be obtained as the absorption probability of the Markov chain described hereafter.

\noindent\textbf{States.} The epidemic can be described as a Markov chain on the following $2+\frac{N(N-1)}{2}$ states:

\begin{itemize}
   \item $Init$: The initial state in which only the origin $a$ is
     infected. This state is transient.
   \item $Succ$: The destination $b$ has been infected. This state is absorbing.
   \item States $(i,j)$ for $1 \le i \le N-1$ and $0 \le j \le   N-1-i$. These are also transient.
\end{itemize}

\noindent\textbf{Primitives.} The transition probabilities are functions of the following primitives. Given two sets of nodes $U$ and $W$, if each node of $U$ can infect each other node in $W$ with probability $p$, we define the probability that $m$ nodes in $W$ will be infected:
\begin{equation}
P_{inf}(m,p,|U|,|W|) = \textrm{pdf}_{\mathcal{B}}\left(m,1-(1-p)^{|U|},|W|\right),
\end{equation}
\noindent where $\textrm{pdf}_{\mathcal{B}}(m,p,n)$ is the probability density function of a binomial distribution of $n$ independent events with probability $p$.

A node that has just been infected (i.e., $\in J_k$) can contaminate the destination the following round with probability $\pi_\uparrow$, while nodes that have been infected for two or more time steps (i.e.,~$\in I_k$) can do so with probability $p_\uparrow$. If $|I_k|=i$ and $|J_k|=j$, then the probability of infecting the destination $b$ during the next time step is:
\begin{equation}
P_{succ}(i,j) = 1 - \pi_\downarrow^j (1-p_\uparrow)^i.
\label{p_succ}
\end{equation}

\noindent\textbf{Transition Probabilities.} The state $Succ$ is absorbing. Any transitions from the $Init$ state, can be calculated as transitions from a $(0,1)$ state. A state $(i,j)$ can transition to either state $Succ$ with probability $P_{succ}(i,j)$ or to another state $(i+j,j')$ with probability:
\begin{eqnarray}
\left( 1-P_{succ}(i,j) \right) \sum_{m=0}^{j'} \Big\{ P_{inf}(m,\pi_\uparrow,j,N-1-i-j) \nonumber \\ 
\times P_{inf}(j'-m,p_\uparrow,i,N-1-i-j-m) \Big\}. \label{p_trans}
\end{eqnarray}

\noindent\textbf{Delivery Ratio.} Let $\mathbf{T}$ be the Markov chain's matrix of transition probabilities, $\mathbf{i}$ the initial state vector and $\mathbf{s}$ the state vector with coefficient $1$ for state $Succ$ and $0$ for all others.  Therefore, the delivery ratio (i.e., the probability of being in state $Succ$) after $d$ time steps is \mbox{$P_{deliv}( d, \alpha = 1 ) = \mathbf{i} \mathbf{T}^d \mathbf{s}$}.

For example, let us consider a network with $3$ mobile nodes. The Markov chain describing an epidemic propagation on its associated edge-Markovian dynamic graph has $5$ states: $Init = (0,1)$, $(1,0)$, $(1,1)$, $(2,0)$, and $Succ$. Here $s = [0 \: 0 \: 0 \: 0 \: 1]^T$ and its initial vector state vector is $i = [1 \: 0 \: 0 \: 0 \: 0]$. Its matrix of transition probabilities, $\mathbf{T}$, is detailed in Eq.~(\ref{eqn:example}).

\subsection{Bundles smaller than link size \texorpdfstring{($\alpha < 1$)}{}}
\label{subsec:p_smaller}

When the bundle size is smaller than the link size, bundles may perform up to $\left \lfloor \frac{1}{\alpha} \right \rfloor$ hops during one time step. Recall that the network topology instantly changes at the beginning of each time step, before the \emph{first} hop. After that, the remaining hops happen on the same \emph{static} network topology. To take this into account, we define a \emph{static} propagation matrix $\mathbf{R}$ using the same states as previously but tweaking the transition probabilities. In a static topology no new links can come up, hence $P_{succ}^{static}(i,j) = 1 - \pi_\downarrow^j$ and the transition from state $(i,j)$ to $(i+j,j')$ happens with probability \mbox{$\left( 1-P_{succ}^{static}(i,j) \right) P_{inf}(j',\pi_\uparrow,j,N-1-i-j)$}. Finally, the delivery ratio (i.e., the probability of being in $Succ$ after $d$ time steps) is \mbox{$P_{deliv}(d,\alpha < 1) = \mathbf{i} \left( \mathbf{T} \cdot \mathbf{R}^{\lfloor \frac{1}{\alpha} \rfloor -1} \right)^d \mathbf{s}$}.

\subsection{Bundles larger than link size \texorpdfstring{($\alpha > 1$)}{}}
\label{subsec:p_greater}

Bundles larger than the link size can only use links that last longer than $\lceil \alpha \rceil$ time steps. Computing the exact delivery ratio in this case requires one to keep track of the number of nodes that will complete reception of the bundle in $1,2, \ldots, \lceil \alpha \rceil$ time steps. This quickly becomes intractable. Instead one can easily calculate upper and lower bounds on the delivery ratio by considering successive, non overlapping, intervals of $\lceil \alpha \rceil$ time steps and only the links that last longer than $\lceil \alpha \rceil$ time steps. The latter will hereafter be referred to as \textit{sufficiently long} links.

The lower bound is obtained by taking into account only the \textit{sufficiently long} links that either exist or come up at the \emph{beginning} of an interval. This ignores links that come up later in the interval and hence underestimates the propagation of the epidemic. More precisely, we replace $\pi_\uparrow$ by $\pi_\uparrow (1-p_\downarrow)^{\lceil \alpha \rceil - 1}$ and $p_\uparrow$ by $p_\uparrow p_\downarrow^{\lceil \alpha \rceil - 1}$ in Eqs.~\ref{p_succ} and \ref{p_trans}.

The upper bound is obtained by considering that \emph{any} \textit{sufficiently long} link that comes up during one interval will allow the \emph{full} bundle to be transmitted over it by the end of the interval. For example, if $\alpha = 2$ and a \textit{sufficiently long} link appears after one time step within the two-time-step interval, then we consider that the whole bundle can be transferred over that link before the next interval. This obviously overestimates the number of infected nodes at each time step. More precisely, we replace $\pi_\uparrow$ by $\left(\pi_\uparrow+\pi_\downarrow (1-(1-p_\uparrow)^{\lceil \alpha \rceil - 1})\right) (1-p_\downarrow)^{\lceil \alpha \rceil - 1}$ and $p_\uparrow$ by $\left(1-(1-p_\uparrow)^{\lceil \alpha \rceil}\right)(1-p_\downarrow)^{\lceil \alpha \rceil - 1}$ in Eqs.~\ref{p_succ} and \ref{p_trans}.

If $\mathbf{T_l}$ (resp. $\mathbf{T_u}$) is the transition matrix obtained for the lower (resp. upper) bound, then the delivery ratio after $d$ time steps is bounded by \mbox{$\mathbf{i} \mathbf{T_l}^{\frac{d}{\lceil \alpha \rceil}} \mathbf{s} \le P_{deliv}(d,\alpha > 1) \le \mathbf{i} \mathbf{T_u}^{\frac{d}{\lceil \alpha \rceil}} \mathbf{s}$}.

\section{Discussion}
\label{sec:discussion}

\subsection{Influence of bundle size}
\label{subsec:bundle_size}

\begin{figure}[t]
  \centering
  \includegraphics{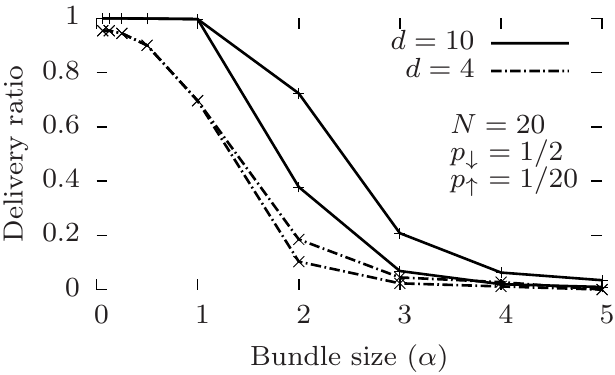}
  \caption{Influence of bundle size on delivery ratio for different values of maximum delay ($d$). Each value of $d$ corresponds to two lines: its upper and lower bounds.}
  \label{param_size}
\end{figure}

Fig.~\ref{param_size} plots the delivery ratio as a function of the bundle size for different values of maximum delay. Bundles larger than the link size see their delivery ratio severely degraded, though this is somewhat mitigated by longer maximum delays. On the other hand, bundles smaller than the link size can make several hops in a single time step. This is a great advantage when the time constraints are particularly tight ($d=4$ in Fig.~\ref{param_size}), but barely has any effect when the time constraints are looser. This also highlights the influence of node mobility. Indeed, since the actual bundle size is proportional to $\tau$ (see Section~\ref{subsec:preliminaries}), high node mobility (i.e., small $\tau$) makes the actual link size smaller and thus further constrains possible bundle size.

\subsection{Influence of other parameters}
\label{subsec:params}

\begin{figure}[t]
  \centering
  \includegraphics{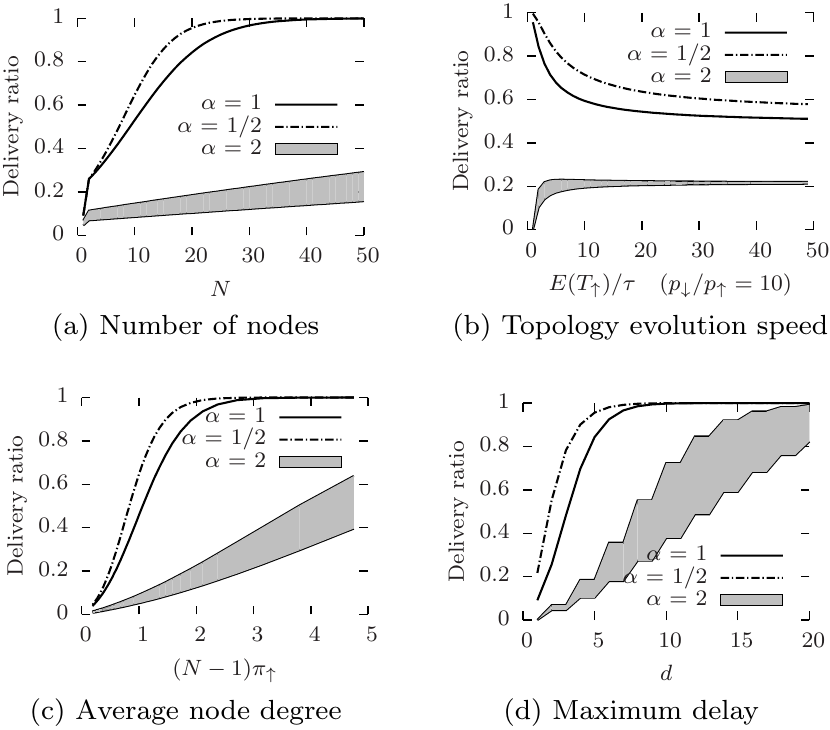}
  \caption{Influence of model parameters on the delivery ratio. When unspecified, $N=20$, $p_\downarrow=1/2$, $p_\uparrow=1/20$, $d=5$. Maximum delay and average link lifetime are expressed in number of time steps.}
  \label{params_influence}
\end{figure}

\noindent\textbf{Number of nodes.}
(Fig.~\ref{params_influence}a) The delivery ratio tends to $1$ as $N$ increases. Indeed, for a given source/destination pair, each new node is a new potential relay in the epidemic dissemination and thus helps the delivery ratio.

\noindent\textbf{Topology evolution speed.}
(Fig.~\ref{params_influence}b) Faster oscillations between $\uparrow$ and $\downarrow$ states make for a more dynamic network topology. This makes for shorter contact and inter-contact times (Section \ref{sec:markovian_graphs}) but increases contact opportunities. Small bundles ($\alpha \le 1$) take advantage of this and their delivery ratio increases as $E(T_\uparrow)$ decreases. On the other hand, excessive link instability drives the delivery ratio for larger bundles ($\alpha > 1$) to 0, because fewer links last longer than one time step.

\noindent\textbf{Average node degree.}
(Fig.~\ref{params_influence}c) Greater connectivity increases the delivery ratio. The sharp slope of the curve when $\alpha \le 1$ is reminiscent of percolation in random graphs when the average node degree hits $1$.

\noindent\textbf{Maximum Delay.}
(Fig.~\ref{params_influence}d) All else being equal, there is a threshold value beyond which almost all bundles are delivered. This can be linked to the space-time diameter of the underlying topology~\cite{chaintreau_diam}.

\section{Evaluation}
\label{sec:experimental}

The theoretical results from the previous section give us valuable insights into real-life scenarios. Although the edge-Markovian model's diameter is significantly smaller than that of real-work networks due to unwanted small-world properties, it accurately predicts, as we shall see in this section, the relations between delivery ratio, maximum delay and bundle size.

\subsection{Methodology}

Wireless connectivity traces involving mobile devices have typically been conducted using periodic Bluetooth scans~\cite{chaintreau_mobility,mit,tournoux08_rollernet}. In this paper, we chose to study the Rollernet trace~\cite{tournoux08_rollernet}, which captures the connectivity patterns in a rollerblading tour, because of its very short sampling period. Indeed, the longer the sampling period, the more likely link failures or short contacts will be missed. Furthermore, it becomes difficult to claim that a contact translates into a link that lasts roughly as long as the sampling period (one of our core theoretical assumptions). Therefore, in order to compare theoretical and experimental results, we require traces with very short sampling periods.

Other Bluetooth contact traces were considered, such as the Reality Mining experiment conducted at MIT~\cite{mit} or the Infocom 2005 traces from the Haggle Project~\cite{chaintreau_mobility}. Unfortunately, none of these had a short enough sampling period (600 and 120 seconds respectively, compared to 15 seconds for Rollernet). In a sense, the MIT and Infocom traces capture a subset of contact opportunities while Rollernet approaches the evolution of the connectivity graph.

Since the dataset logs contacts between nodes and not link durations, we assumed that two nodes in contact remain so for the entire sampling period. Furthermore, we did not try to extrapolate additional events (e.g., new contact opportunities and link failures) between multiples of the sampling period. As in the Markovian network model described previously, we again assume that all links have equal capacity. The first 3,000 seconds of Rollernet trace were replayed. Every 15 seconds for the first 2,000 seconds, 60 source/destination pairs were randomly selected for a simulation of epidemic routing. The average link lifetime is $26.18$ seconds and the average node degree $4.75$. Using the expressions from Section~\ref{sec:markovian_graphs}, we derive $p_\uparrow=0.05$ and $p_\downarrow=0.57$.

\subsection{Results}

\begin{figure}[t]
  \centering
  \includegraphics{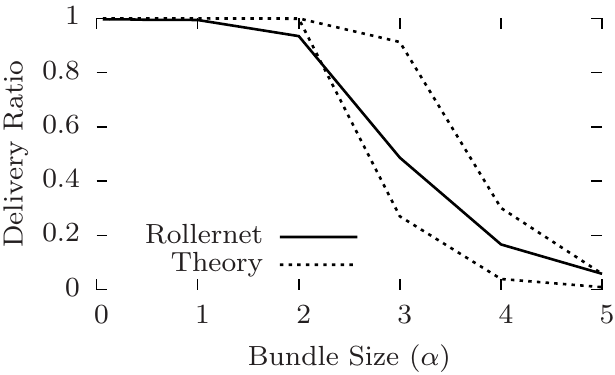}
  \caption{Predicting delivery ratio for different bundle sizes in Rollernet with a 5-minute maximum delay.}
  \label{exp_theory}
\end{figure}

\begin{figure}[t]
  \centering
  \includegraphics{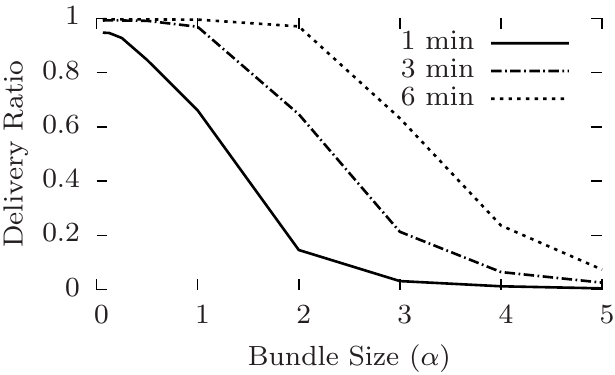}
  \caption{Rollernet: Delivery ratio vs. bundle size for various maximum delay values.}
  \label{exp_results}
\end{figure}

\noindent\textbf{Predicting delivery ratio.} Fig.~\ref{exp_theory} compares the measured delivery ratio in Rollernet to the model's upper and lower bounds. Here we use the $p_\uparrow$ and $p_\downarrow$ values derived from the trace's average link lifetime and node degree. Due to small-world effects, our model is overly optimistic, particularly for smaller bundle sizes ($\alpha \le 2$). However it does successfully bound the experimental values for larger bundle sizes.

\noindent\textbf{Smaller bundles increase delivery ratio.} In Fig.~\ref{exp_results}, the delivery ratio is steady and close to $1$ before dropping sharply beyond a certain bundle size that depends on the target delay. Due to mobility, more than half of the links last less than $15$ seconds. Therefore, bundles of size greater than $1$ forgo many contact opportunities. However, longer maximum delays can compensate for this. This mirrors the theoretical results on size, delay, and mobility described in Section.~\ref{subsec:bundle_size}.

\noindent\textbf{Bounded gain from smaller bundles.} In Fig.~\ref{exp_results}, when the maximum delay is $1$ minute, the maximum achievable delivery ratio is $0.95$ no matter how small the bundles are. This bound on the gain achieved by smaller bundles appears because they hit the performance limit of epidemic routing. Indeed, the best possible epidemic propagation of a message will, at each time step, infect a whole connected component if at least one of its nodes is infected. A small enough bundle can spread sufficiently quickly to achieve this, and thus even smaller bundles bring no performance gain. The same bounded gain from smaller bundles is visible on Fig.~\ref{param_size} on the $d=4$ curve.

\noindent\textbf{Tight delays require smaller bundles.} The sharp delivery ratio drop in Fig.~\ref{exp_results} occurs later for more relaxed delay constraints. A tight time constraint (less than a couple of minutes for example) forces the use of smaller bundles in order to obtain an acceptable delivery ratio. On the other hand, looser time constraints allow for more flexibility regarding bundle size. It is therefore possible to determine the maximum bundle size for any given target delivery ratio. 

\vfill \break

\section{Conclusion}
In this paper, we proposed a new model for epidemic propagation on edge-Markovian dynamic graphs which capture the correlation between successive connectivity graphs. We find a closed-form expression of delivery ratio as a function of bundle size, maximum tolerated delay, and the dynamics of the underlying evolving graph. In particular, we have shown that, given a certain maximum delay and node mobility, bundle size has a major impact on the delivery ratio. Our theoretical insights on the interaction between these parameters are corroborated by experimental results on the Rollernet dataset.

\bibliographystyle{IEEEtran} 
\bibliography{whitbecktoc}

\end{document}